\documentclass[longauth]{aa} 
\usepackage{graphicx} \usepackage{lscape} \usepackage{longtable}
\usepackage{txfonts}
\usepackage{longtable,lscape}
\usepackage{rotating} 
\usepackage{natbib}
\usepackage[colorlinks=true,linkcolor=blue,citecolor=blue]{hyperref}

\newcommand{\mjup}{M$_{\rm Jup}$}

 \newcommand{\bp}{$\beta$\,Pictoris\,} \newcommand{\bpic}{$\beta$\,Pictoris\,}
  
\renewcommand{\muup}{$\mu$m}

\begin{document}

   \title{  Post-conjunction detection of \bpic b with VLT/SPHERE \thanks{Based on observations collected at the European Southern Observatory under programmes 198.C-0209, 1100.C-0481}}
  %
  
\author{
 A.-M. Lagrange\inst{\ref{ipag}}
\and A. Boccaletti\inst{\ref{lesia}}
\and M. Langlois\inst{\ref{lam},\ref{cral}}
\and G. Chauvin\inst{\ref{ipag},\ref{umi}}
\and R. Gratton\inst{\ref{padova}}
\and H. Beust\inst{\ref{ipag}}
\and S. Desidera \inst{\ref{padova}}
\and J. Milli \inst{\ref{eso}}
\and M. Bonnefoy\inst{\ref{ipag}}
\and A. Cheetham\inst{\ref{geneva}}
\and M. Feldt\inst{\ref{mpia}}
\and M. Meyer \inst{\ref{eth},\ref{michigan}}
\and A. Vigan\inst{\ref{lam}}
\and B. Biller\inst{\ref{mpia},\ref{edinburgh}}
\and M. Bonavita\inst{\ref{padova},\ref{edinburgh}}
\and J.-L. Baudino\inst{\ref{padova},\ref{oxford}} 
\and F. Cantalloube\inst{\ref{mpia}}
\and M. Cudel\inst{\ref{ipag}}
\and S. Daemgen \inst{\ref{eth}}
\and P. Delorme \inst{\ref{ipag}}
\and V. D'Orazi\inst{\ref{padova}} 
\and J. Girard\inst{\ref{ipag}}
\and C. Fontanive\inst{\ref{padova},\ref{edinburgh}}
\and J. Hagelberg\inst{\ref{ipag}}
\and M. Janson \inst{\ref{mpia},\ref{stockholm}}
\and M. Keppler \inst{\ref{mpia}}
\and T. Koypitova \inst{\ref{mpia}}
\and R. Galicher \inst{\ref{lesia}}
\and J. Lannier \inst{\ref{ipag}}
\and H. Le Coroller\inst{\ref{lam}}
\and R. Ligi \inst{\ref{lam},\ref{merate}}
\and A.-L. Maire \inst{\ref{mpia}}
\and D. Mesa \inst{\ref{padova}}
\and S. Messina\inst{\ref{catania}}
\and A. M\"ueller\inst{\ref{mpia}}
\and S. Peretti\inst{\ref{geneva}} 
\and C. Perrot\inst{\ref{lesia}} 
\and D. Rouan \inst{\ref{lesia}}
\and G. Salter \inst{\ref{lam}}
\and M. Samland\inst{\ref{mpia}}
\and T. Schmidt\inst{\ref{lesia}}
\and E. Sissa\inst{\ref{padova}}
\and A. Zurlo\inst{\ref{lam},\ref{nucleo}}
\and J.-L. Beuzit\inst{\ref{ipag}} 
\and D. Mouillet\inst{\ref{ipag}}
\and C. Dominik\inst{\ref{amsterdam}}
\and T. Henning\inst{\ref{mpia}}
\and E. Lagadec\inst{\ref{oca}}
\and F. M\'enard\inst{\ref{ipag}}
\and H.-M. Schmid\inst{\ref{eth}}
\and M. Turatto\inst{\ref{padova}}
\and S. Udry\inst{\ref{geneva}}  
\and A.J. Bohn\inst{\ref{leiden}}
\and B. Charnay\inst{\ref{lesia}}
\and C. A. Gomez Gonzales\inst{\ref{ipag}}
\and C. Gry\inst{\ref{lam}}
\and M. Kenworthy\inst{\ref{leiden}}
\and Q. Kral\inst{\ref{lesia}}
\and C. Mordasini\inst{\ref{bern}}
\and C. Moutou\inst{\ref{lam}}
\and G. van der Plas\inst{\ref{leiden}}
\and J. E. Schlieder\inst{\ref{goddard},\ref{geneva}}
\and  L. Abe\inst{\ref{oca}}
\and  J. Antichi\inst{\ref{firenze}}
\and  A. Baruffolo\inst{\ref{padova}}
\and  P. Baudoz\inst{\ref{lesia}}
\and  J. Baudrand\inst{\ref{lesia}}
\and  P. Blanchard\inst{\ref{lam}} 
\and  A. Bazzon\inst{\ref{eth}}
\and  T. Buey\inst{\ref{lesia}}
\and  M. Carbillet\inst{\ref{oca}}
\and  M. Carle\inst{\ref{lam}} 
\and  J. Charton\inst{\ref{ipag}}
\and  E. Cascone\inst{\ref{napoli}}
\and  R. Claudi\inst{\ref{padova}}
\and  A. Costille\inst{\ref{lam}} 
\and  A. Deboulbe\inst{\ref{ipag}}
\and  V. De Caprio\inst{\ref{napoli}}
\and  K. Dohlen\inst{\ref{lam}} 
\and  D. Fantinel\inst{\ref{padova}}
\and  P. Feautrier\inst{\ref{ipag}}
\and  T. Fusco\inst{\ref{onera}}
\and  P. Gigan\inst{\ref{lesia}} 
\and  E. Giro\inst{\ref{padova}}
\and  D. Gisler\inst{\ref{eth}}
\and  L. Gluck\inst{\ref{ipag}}
\and  N. Hubin\inst{\ref{garching}}
\and  E. Hugot\inst{\ref{lam}}
\and  M. Jaquet\inst{\ref{lam}}
\and  M. Kasper\inst{\ref{garching}}
\and  F. Madec\inst{\ref{lam}}
\and  Y. Magnard\inst{\ref{ipag}}
\and  P. Martinez\inst{\ref{oca}}
\and  D. Maurel\inst{\ref{ipag}}
\and  D. Le Mignant\inst{\ref{lam}}
\and  O. M\"oller-Nilsson\inst{\ref{mpia}}
\and  M. Llored\inst{\ref{lam}}
\and T. Moulin\inst{\ref{ipag}}
\and  A. Orign\'e\inst{\ref{lam}}
\and A. Pavlov\inst{\ref{mpia}}
\and D. Perret\inst{\ref{lesia}}
\and C. Petit\inst{\ref{onera}}
\and J. Pragt\inst{\ref{nova}}
\and J. Szulagyi\inst{\ref{zurich}}
\and F. Wildi\inst{\ref{geneva}}
}

\institute{
Univ. Grenoble Alpes, CNRS, IPAG, F-38000 Grenoble, France\label{ipag}
\and LESIA, Observatoire de Paris, PSL Research University, CNRS, Sorbonne Universités, UPMC Univ. Paris 06, Univ. Paris Diderot, Sorbonne Paris Cité, 5 place Jules Janssen, 92195 Meudon, France\label{lesia}
\and Aix Marseille Universit\'e, CNRS, LAM (Laboratoire d'Astrophysique de Marseille) UMR 7326, 13388 Marseille, France\label{lam}
\and CRAL, UMR 5574, CNRS, Universit de Lyon, Ecole Normale Supri\'eure de Lyon, 46 All\'ee d'Italie, F-69364 Lyon Cedex 07, France\label{cral}
\and Unidad Mixta Internacional Franco-Chilena de Astronom\'{i}a, CNRS/INSU UMI 3386 and Departamento de Astronom\'{i}a, Universidad de Chile, Casilla 36-D, Santiago, Chile\label{umi}
\and INAF - Osservatorio Astronomico di Padova, Vicolo dell’ Osservatorio 5, 35122, Padova, Italy\label{padova}
\and ESO Alonso de Córdova 3107, Vitacura, Región Metropolitana, Chili\label{eso}
\and Max Planck Institute for Astronomy, K\"onigstuhl 17, D-69117 Heidelberg, Germany\label{mpia}
\and Instiute for Particle Physics and Astrophysics, ETH Zurich, Wolfgang-Pauli-Strasse 27, 8093 Zurich, Switzerland\label{eth}
\and Department of Astronomy, University of Michigan, Ann Arbor, MI 48109, US\label{michigan}
\and SUPA, Institute for Astronomy, The University of Edinburgh, Royal Observatory, Blackford Hill, Edinburgh, EH9 3HJ, UK\label{edinburgh}
\and Department of Astrophysics, Denys Wilkinson Building, Keble Road, Oxford, OX1 3RH, UK\label{oxford}
\and Department of Astronomy, Stockholm University, AlbaNova University Center, SE-10691, Stockholm, Sweden\label{stockholm}
\and INAF-Osservatorio Astrofisico di Catania, Via S. Sofia 78, I-95123 Catania, Italy\label{catania}
\and Geneva Observatory, University of Geneva, Chemin des Mailettes 51, 1290 Versoix, Switzerland\label{geneva}
\and N\'ucleo de Astronomía, Facultad de Ingenier\'ia, Universidad Diego Portales, Av. Ejercito 441, Santiago, Chile\label{nucleo}
\and Anton Pannekoek Institute for Astronomy, Science Park 904, NL-1098 XH Amsterdam, The Netherlands\label{amsterdam}
\and Universit\'e Cote d’Azur, OCA, CNRS, Lagrange, France\label{oca}
\and Leiden Observatory, Leiden University, PO Box 9513, 2300 RA Leiden, The Netherlands\label{leiden}
\and INAF-Osservatorio Astronomico di Brera, Via E. Bianchi 46, I-23807 Merate, Italy\label{merate}
\and Exoplanets and Stellar Astrophysics Laboratory, Code 667, NASA Goddard Space Flight Center, Greenbelt MD, 20771, USA\label{goddard}
\and Physikalisches Institut, Universität Bern, Gesellschaftsstrasse 6, 3012, Bern, Switzerland\label{bern}
\and INAF - Osservatorio Astrofisico di Arcetri, Largo E. Fermi 5, I-50125 Firenze, Italy\label{firenze}
\and INAF - Osservatorio Astronomico di Capodimonte, Salita Moiariello 16, 80131 Napoli, Italy\label{napoli}
\and ONERA (Office National d’Etudes et de Recherches A\'erospatiales), B.P.72, F-92322 Chatillon, France\label{onera}
\and European Southern Observatory (ESO), Karl-Schwarzschild-Str. 2, 85748 Garching, Germany\label{garching}
\and NOVA Optical Infrared Instrumentation Group, Oude Hoogeveensedijk 4, 7991 PD Dwingeloo, The Netherlands\label{nova}
\and Center for Theoretical Astrophysics and Cosmology, Inst. for Computational Science, University of Z\"urich, Winterthurerstrasse 190, CH-8057 Z\"urich, Switzerland\label{zurich}
}
   \date{Received date: this version XXX / Accepted date}

   %
   
\abstract
    {With an orbital distance comparable to that of Saturn in the solar system, \bpic  b  is the closest (semi-major axis $\simeq$\,9\,au) exoplanet that has been imaged to orbit a star. Thus it offers unique opportunities for detailed studies of its orbital, physical, and atmospheric properties, and of disk-planet interactions. With the exception of the discovery observations in 2003 with NaCo at the Very Large Telescope (VLT), all following astrometric measurements relative to \bpic have been  obtained in the southwestern part of the orbit, which severely limits the determination of the planet's orbital parameters.}
    {We aimed at further constraining \bpic b orbital properties using more data, and, in particular, data taken in the northeastern part of the orbit.}
       {We used  SPHERE  at the VLT to precisely monitor the orbital motion of beta \bpic  b since first light of the instrument in 2014.}
    { We were able to monitor the planet until November 2016, when its angular separation became too small (125 mas, i.e., 1.6\,au) and prevented further detection.  We  redetected \bpic b on the northeast side of the disk at a  separation of 139\,mas and a PA of 30$^{\circ}$ in September 2018. The  planetary orbit is now well constrained. With a semi-major axis (sma)  of $a = 9.0  \pm 0.5$ au (1 $\sigma $), it  definitely excludes previously reported possible long orbital periods, and excludes \bpic b as the origin of photometric variations that took place in 1981. We also refine the eccentricity and inclination of the planet. From an instrumental point of view, these data demonstrate that it is possible to detect, if they exist, young massive Jupiters  that  orbit at less than 2 au from a star that is 20 pc away.}

 \offprints{Anne-Marie Lagrange, \email{anne-marie.lagrange 'at' univ-grenoble-alpes.fr} }

\titlerunning{}

\authorrunning{A.-M. Lagrange et al.}

\maketitle
%
%
\section{Introduction}

With its imaged debris disk of dust \cite[see ][ for the discovery image]{smith84}, its falling, evaporating exocomets \cite[][and references therein]{kiefer14}, and an imaged giant planet \cite[][]{lagrange10}, the $\sim $ 20 Myr old \bpic is a unique proxy for the study of the early stages of planetary system formation and evolution, when giant planets are formed, Earth-mass planets may still be forming, and most of the protoplanetary gas has disappeared from the disk. Its proximity to Earth \cite[][distance = $19.454\pm 0.05$ pc]{vanLeeuwen07} and the  relatively short ($\simeq$ two decades) orbital period of \bpic b enable detailed studies of its orbit and its physical and atmospheric properties. The system also allows us to study the interaction between planet(s) and disks. \bpic b can explain, for example, several (but not all) of the dust disk morphologies, in particular its inner warp, and some outer asymmetries \cite[][]{lagrange10}. \bpic b could also be the trigger for the infall and evaporation of cometary bodies (exocomets) onto the star, if it has a non-zero eccentricity \cite[][]{bmo96}. Last, \bpic b was suggested to be responsible for the photometric variations observed in 1981 \cite[][]{leca95} and has tentatively been attributed to a planet transit \cite[][]{leca97}.

Careful monitoring of the position of the planet relative to the star (referred to as astrometric measurements) with the
Nasmyth Adaptive Optics System (NAOS) Near-Infrared Imager and Spectrograph (CONICA)
 (NaCo) at the Very Large Telescope (VLT) constrained its orbital properties \cite[][]{bonnefoy14,chauvin12}. By combining data from NaCo and the Gemini Planetary Imager (GPI), \cite{wang16} found a similar bet slightly different orbit. Assuming that the peculiar photometric event observed in 1981 could be due to the transit of \bpic b  in front of the star, \cite{leca16} identified a second family of orbital solutions: a semi-major axis a = 13 au instead of 9 au, that is, a period of 34 yr instead of about 20 yr, and e = 0.3 instead of less than 0.1. Recently, \cite{snellen18} used GAIA\ and HIPPARCOS measurements to constrain the planet period to $\geq$ 22 yr, and reported a mass of $11\pm2$ \mjup, which is compatible with the constraints derived from radial velocity data alone \cite[][]{lagrange12rv} or by the combined analysis of direct imaging and radial velocity data  \cite[][]{bonnefoy14}.

\begin{table*}
\caption{Observing log}
\label{logobs}
\begin{center}
\renewcommand{\footnoterule}{} 
\small
\begin{tabular}{lllllllllll}
\hline \hline
Date UT       & Filters      
& DIT$\times$NDIT$\times$Nexp & AM      & $\Delta$par  & DIMM seeing  & $\tau$$_ 0$  & TNcorr  & Plate scale \\
 (yyyy-mm-dd)     &        &(s)            
 &    & ($^{\circ}$)    & ($''$)        & (ms)  & ($^{\circ}$) & (mas/pix) \\
\hline
2014-12-08  & IRDIS-K1K2    & 4$\times$40$\times$36    & 1.12       & 16.4  
         & 0.75          &  4.4          & -1.71         & 12.251 \\

%
2015-02-05  & IRDIS-H2H3    & 4 $\times$40$\times$8    & 1.16         & 43.0    & 0.78  & 7.8  & -1.72  & 12.255 \\
%
2015-10-01  & IRDIS-H2H3 & 4$\times$16 $\times$2      & 1.15       & 22.8    & 1.42  & 1.0  & -1.81 & 12.250 \\
%
2015-11-30  & IRDIS-H2H3 & 4$\times$80$\times$32      & 1.12        & 39.9    & 1.06  & 10.1  & -1.75 & 12.255 \\
%
2015-12-26  & IRDIS-H2H3 & 8$\times$98$\times$20      & 1.12        & 36.6    & 1.44  & 2.1  & -1.79 & 12.255 \\
%
2016-01-20  & IRDIS-H2H3 & 4$\times$60$\times$30      & 1.12       & 29.3    & 0.60  & 3.0  & -1.81 & 12.255 \\
%
2016-03-27  & IRDIS-H2H3 & 8$\times$20$\times$32      & 1.28        & 19.9    & 0.74  & 1.7  & -1.77 & 12.255 \\
%
2016-04-16  & IRDIS-H2H3 & 8$\times$42$\times$32      & 1.42        & 17.9    & 0.65  & 6.0  & -1.74 & 12.255 \\

2016-09-16  & IRDIS-H2H3 & 8$\times$16$\times$8      & 1.17       & 38.6    & 0.32  & 1.2  & -1.76 & 12.255 \\
%
2016-10-14  & IRDIS-H2H3 & 16$\times$8$\times$10      & 1.12        & 53.8    & 0.75  & 3.0   & -1.76 & 12.255 \\
%
2016-11-18  & IRDIS-H2H3 & 2$\times$64$\times$50      & 1.12        & 39.8    & 0.96  & 2.0  & -1.76 & 12.248 \\

2018-09-17  & IRDIS-H2H3  & 2$\times$30$\times$46    & 1.17      & 36.6    & 0.69  & 3.7  & -1.79  & 12.239\\
\hline\hline
\end{tabular}
\end{center}
\tablefoot{AM stands for the mean airmass, $\Delta$par for the variation in parallactic angle during the coronagraphic sequence, TNcorr for true north correction (TN is the angle between the north position and the detector "North"). Seeing and coherence time ($\tau$$_ 0$) are mean values throughout the coronagraphic sequence. }

\end{table*}

\bpic b  was discovered northeast of the star in data obtained in 2003 with NaCo. All available images in addition to this discovery image were obtained in 2009 and later, as the planet orbited  southwest of the star, after it passed behind the star \cite[][]{lagrange10}. The orbital plane is close to equatorial, and the disk of \bpic is seen almost edge-on. This geometrical configuration prevented following the planet in direct imaging when it was projected too close (typically less than 120\,mas) to the star. Altogether, only about 50\% of the planetary orbit had been monitored thus far, including the 2003 epoch, which suffers from relatively large uncertainties and thus limits the precision on the orbital parameters of the planet.  We here present a homogeneous set of planet observations obtained with the Spectro-Polarimetric High-Contrast Exoplanet Research instrument (SPHERE), and in particular, the recent recovery of the planet in September 2018. The observations are described in Section 2, and the results are shown and discussed in Sections 3.

\section{Observations} \label{}

High-contrast coronagraphic SPHERE \citep{beuzit08} observations were obtained between December 2014 and November 2016, and in September 2018, using the IRDIFS mode in the context of the SpHere INfrared survey for Exoplanets (SHINE, \citealt{chauvin17b}). In this setup, the IRDIS \citep{dohlen08} and IFS \citep{claudi08} instruments operate simultaneously. The data were obtained under various atmospheric conditions (see Table\,\ref{logobs}) with the H2 ($\lambda_{c}$= 1.593 \muup ; $\Delta \lambda$ = 0.055 \muup) and H3 ($\lambda_{c}$= 1.667 \muup ; $\Delta \lambda$= 0.056 \muup) narrow bands of IRDIS, except in December 2014, when we used the  K1 ($\lambda_{c}$= 2.1025 \muup ; $\Delta \lambda$ = 0.102 \muup) and K2 ($\lambda_{c}$= 2.255 \muup ; $\Delta \lambda$= 0.109 \muup) narrow bands. The IRDIS images have a field of view (FoV) of $\sim10''\times11''$, and a pixel size of approximately 12.25\,mas. IFS data were also recorded, but they are not analyzed here. We used apodized Lyot coronagraphs that include either a 185\,mas diameter focal mask (N\_ALC\_YJH\_S) or, when the planet was closer to the star, a smaller (145\,mas) mask (N\_ALC\_YJ\_S), combined to an apodizer as well as a pupil stop 
\citep{Carbillet2011}. All coronagraphic data were recorded in stabilized  pupil mode so as to apply angular differential imaging (ADI) post-processing techniques to remove the stellar halo, as described in \cite{marois06}. Most coronagraphic data were also recorded while four satellite footprints of the point spread function (PSF) had been created by the deformable mirror of the instrument and were used for fine monitoring of the frame centering and for photometric purposes. The FoV rotation  during the coronagraphic observations varied between 16$^{\circ}$ and 54$^{\circ}$ (see Table \ref{logobs}). 

Each observing sequence was obtained with the following pattern: PSF - coronagraphic observations - PSF - sky. The PSF data correspond to non-saturated exposures of the star placed out of the coronagraphic mask and obtained using a neutral density filter. They are used for relative photometric reference and to estimate the image quality at the beginning and end of the observations. The sky data were recorded at the end of the coronagraphic sequence to estimate the background level and hot pixels in the science images. Finally, an astrometric field, either Orion or 47 Tuc, was observed with IRDIS for each run \cite[see][]{maire16}. In the case of Orion, the sub-field we used was chosen to be part of the one considered with NaCo since 2008 to allow the best match between the astrometric calibrations of NaCo and SPHERE. The pixel scales and north positions are provided in Table \ref{logobs}.

\begin{figure*}[t!] 
\includegraphics[width=\textwidth]{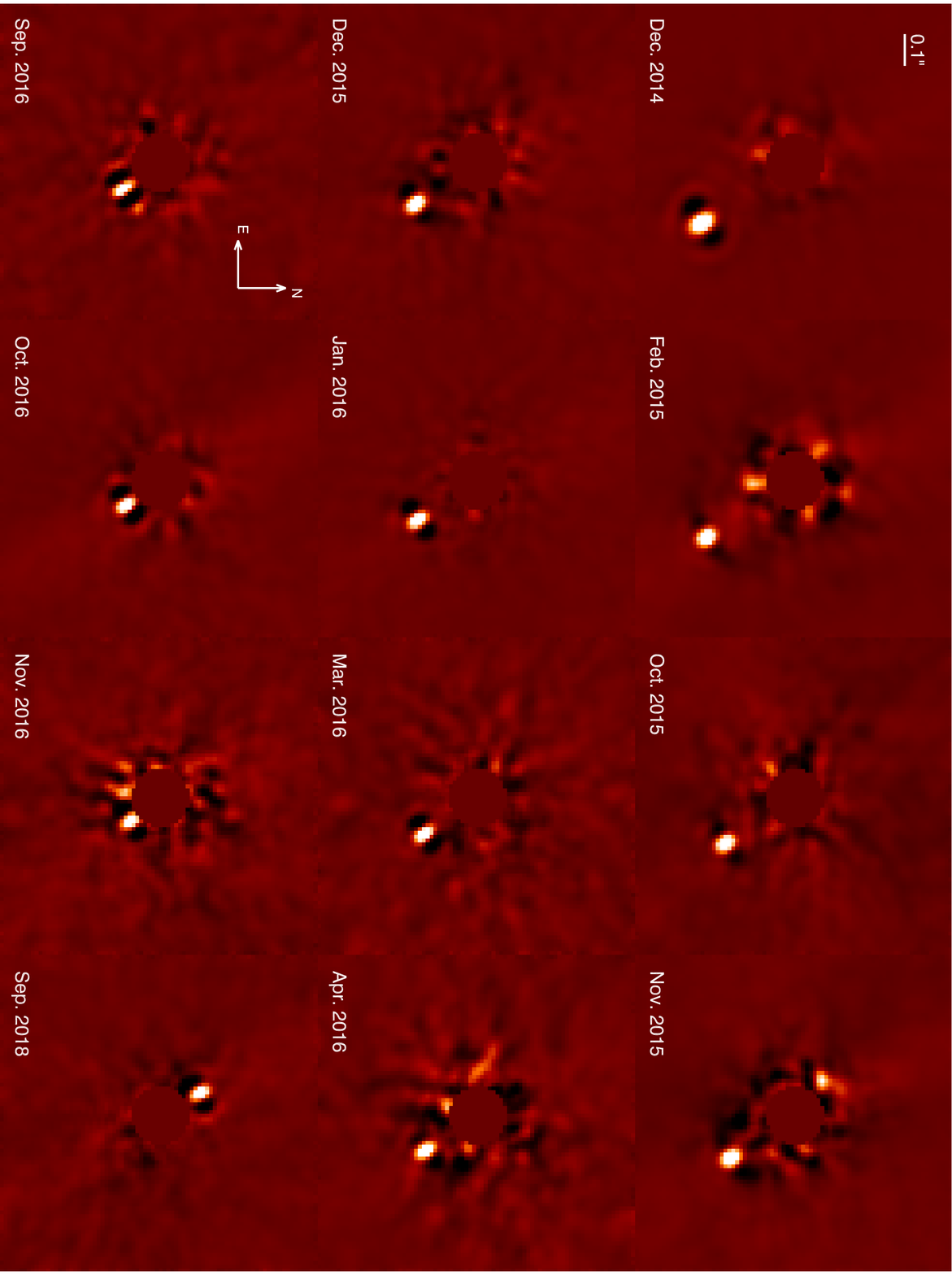} 
\caption{Images of \bpic b with SPHERE IRDIS from December 2014 to September 2018. Each panel displays a FoV of $1''\times1''$. North is up and east is to the left. The intensity scale is adapted at each epoch according to the intensity peak of the planet.} 
\label{images} 
\end{figure*}


The data were reduced as described in \cite{chauvin17} and using the SpeCal tool developed for SPHERE \citep{galicher18}. \bpic b is clearly detected in all images taken between 2014 and 2016, orbiting SW of the star at signal-to-noise ratio (S/N) higher than 9 (Table\,\ref{astrometry}). Fig.\,\ref{images} shows images of the planet at various dates, and Table\,\ref{astrometry} provides the relative position of the planet with respect to the star\footnote{The planet was hardly or even not at all detected in November 2016 with the TLOCI algorithm \citep{Marois14}, while it was detected using a principal component analysis (PCA \cite{soummer12}) algorithm. For consistency, all position measurements were obtained on PCA images.}. The S/N is relatively poor in the last observation of November 2016, as the projected separation of the planet from the star is less than 125 mas ($\simeq $ 1.5 au only) and the contrast is about 9.5 mag, which leads to larger error bars on its astrometry. To our knowledge, neither \bpic b nor any other planet has ever been imaged at such a close projected separation to the star. The 2018 data clearly reveal the planet at 139 mas NE (P.A. about 30$^{\circ}$) from the star.

\section{ Orbital properties of \bpic b}
\begin{figure}[t!] 
\includegraphics[width=\hsize,angle=0]{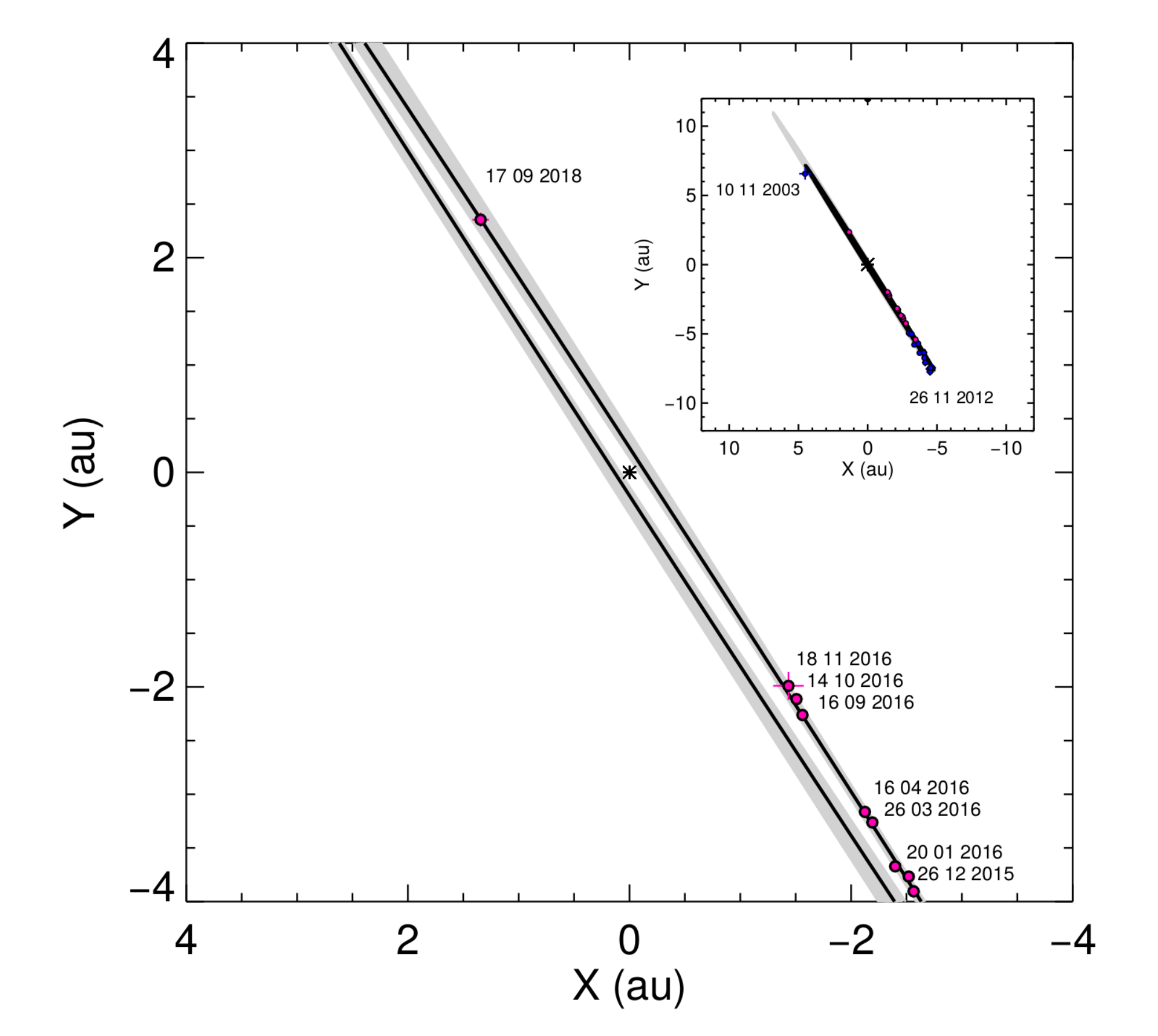} 
\caption{NaCo (\textit{blue}) and SPHERE (\textit{magenta}) astrometric data points of \bpic b shown together with 200 probable solutions of the MCMC analysis  (\textit{gray}) and the best-fit Levenberg-Marquardt solution (\textit{black}). } 
\label{sketch} 
\end{figure}

\begin{figure*}[t!] 
\includegraphics[width=\hsize,angle=0]{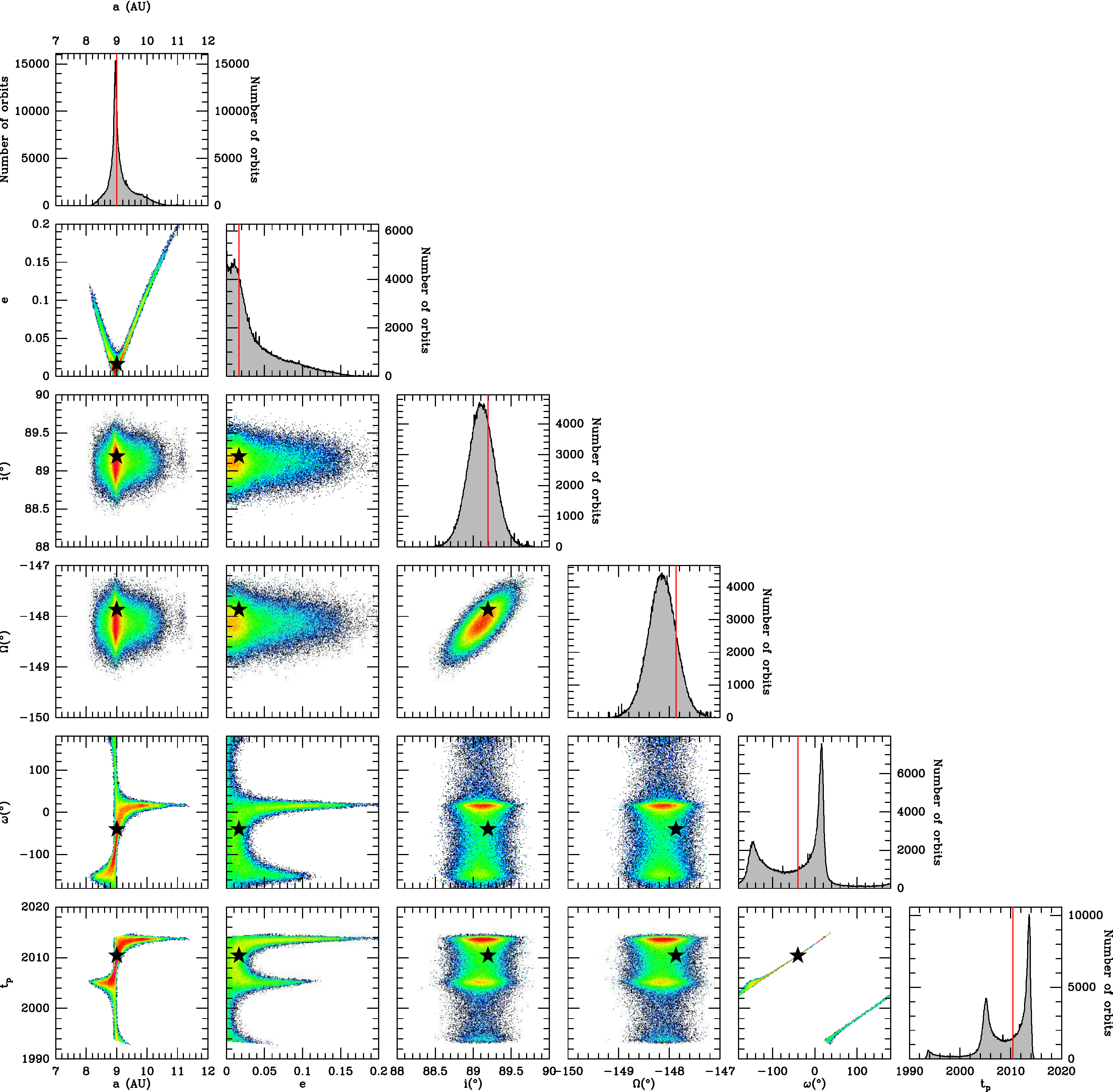} 
\caption{Results of the MCMC fit of the NaCo + SPHERE astrometric data points. The star and the red dotted line correspond to the best-fit solution (best $\chi$2) obtained with a   Levenberg-Marquardt fit.} 
\label{mcmc} 
\end{figure*}

\begin{figure*}[t!] 
\includegraphics[width=\hsize,angle=0]{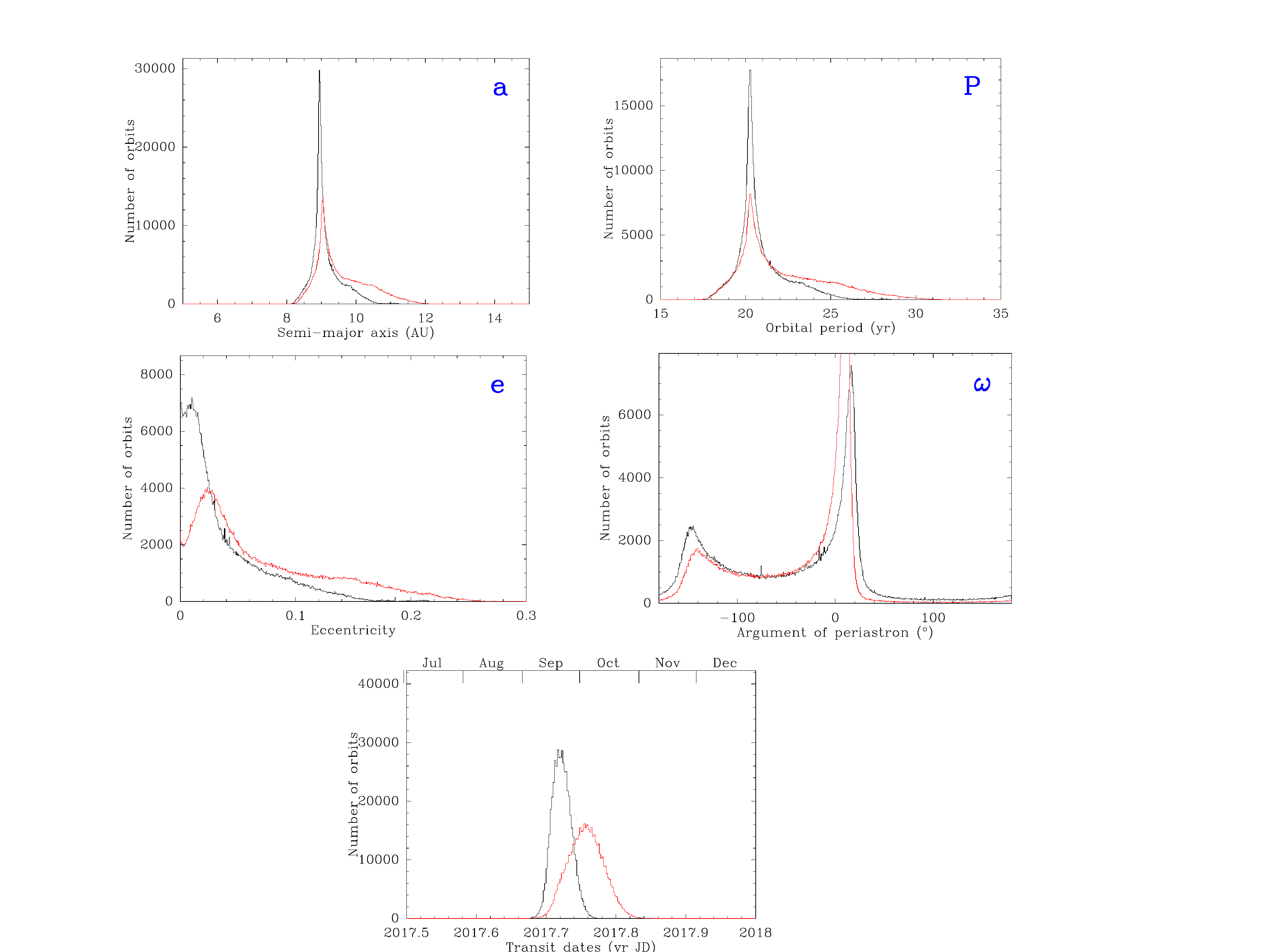} 
\caption{Comparison between the orbital parameters obtained with (black) or without  (red) the recovery point of September 2018.} 
\label{comp} 
\end{figure*}

With these additional SPHERE data, a large part of the orbit is now sampled, as seen in  Fig.\ref{sketch}. We used the positions of the planet relative to the star in these data as well as in the previously published NaCo data \cite[][]{bonnefoy14} together with the Markov chain Monte Carlo  (MCMC)  Bayesian  analysis  technique described in \cite{chauvin12} to derive the probabilistic  distribution  of the orbital  solutions. The results are shown in Fig.\,\ref{mcmc}. For comparison, we show in Fig.\,\ref{comp} the parameters that we deduce with and without the September 2018 data to illustrate the importance of this post-conjunction detection for constraining the orbital properties of the planet.

\begin{table}
\caption{Relative astrometry of \bpic b}
\label{astrometry}
\begin{center}
\renewcommand{\footnoterule}{} 
\small
\begin{tabular}{llll}
\hline \hline
Date UT   & separation (mas)    
& PA ($^{\circ}$)    &  S/N    \\
\hline
2014-12-08 &         $350.51\pm3.20$         & $212.60\pm0.66$    
     &    62  \\
2015-02-05    &        $332.42\pm1.70$         & $212.58\pm0.35$        
&    63\\
2015-10-01    &        $262.02\pm1.78$         & $213.02\pm0.48$        
&    41 \\
2015-11-30    &        $242.05\pm2.51$         & $213.30\pm0.74$        
&    40\\
2015-12-26    &        $234.84\pm1.80$         & $213.79\pm0.51$     
     &    41\\
2016-01-20    &        $227.23\pm1.55$         & $213.15\pm0.46$        
&    62\\
2016-03-26    &        $203.66\pm1.42$         & $213.90\pm0.46$        
&    68\\
2016-04-16    &        $197.49\pm2.36$         & $213.88\pm0.83$        
&    34\\
2016-09-16    &        $142.36\pm2.34$         & $214.62\pm1.10$        
&    18\\
2016-10-14    &        $134.50\pm2.46$         & $215.50\pm1.22$        
&    27\\
2016-11-18    &         $127.12\pm6.44$          & $215.80\pm3.37$   
     &    10\\
2018-09-17    &         $140.46\pm3.12$         & $29.71\pm1.67$        
&    19\\\hline\hline
\end{tabular}
\end{center}
\end{table}

From Fig. 4, we derive the following $1\sigma$ confidence intervals for
the major orbital parameters: $a=8.90^{+0.25}_{-0.41}\,$au for the
semi-major axis, $P=20.29^{+0.86}_{-1.35}$\,yr for the orbital period,
$e=0.01^{+0.029}_{-0.01}$ for the eccentricity, and
$i=89.08^{+0.16}_{-0.19}$ for the inclination. We note that the
orbital period distribution is compatible with a period greater than 22\,yr found by \citet{snellen18} within only $2\sigma $ error bars.

The sma found is compatible with the value derived by \cite{bonnefoy14}, and definitely excludes the long-period solution proposed by \cite{leca16} that assumed a transit of the planet in 1981. \cite{snellen18} predicted a period longer than 22 yr, while the period we derive here is rather 20 yr. We note, however, that the result from the \cite{snellen18} relies on the assumption that there is only one planet around \bpic, which is not necessarily correct. Finally, based on our values, the 2017 conjunction took place in $2017.72\pm 0.04 $, and the next conjunction will occur in 2038.06.


 The orbital solution favors a very low eccentricity but still
remains compatible with zero. We note that the peak of the
distribution is now slightly off zero, in contrast with previous
determinations \cite[][]{mbo14}. Previous studies (\cite{bmo96}, \cite{bmo00} and \cite{thebault01})
of the falling evaporating bodies scenario showed that this phenomenon
could be explained by the perturbing action of a giant planet orbiting
at $\sim 10\,$au from the star, if it has a low orbital
eccentricity. $\beta\:$Pictoris b might of course be assimilated to this
planet, and the fact that its eccentricity might be non-zero is
a key point for this scenario.

Finally, the large FoV of IRDIS enables detecting the disk and the planet in the same image. The SPHERE data confirm the conclusions reached in \cite{lagrange12disk}: the planet projection is between the main disk and the warp. 

\section{Concluding remarks and perspectives} 

The sensitivity of SPHERE allowed us to follow \bpic b down to 125 mas from the star in projected separation. The latest measurements reveal the planet on the NE side of the star, for the first time since its discovery. It was last detected in November 2016 and was redetected in September 2018. Based on the observed data, the semi-major axis of the planet is well constrained to $8.90^{+0.25}_{-0.41}\,$au (1 $\sigma $), its eccentricity to $e=0.01^{+0.029}_{-0.01}$, and its inclination to $89.08^{+0.16}_{-0.19}$ degrees. The data do not support that the planet was  responsible for the photometric event in 1981. More data obtained in the NE part of the disk will allow further refining the orbital properties of the planet. Further work will include the combination of these new, crucial astrometric data points with our latest radial velocity measurements and, if possible, a combination with GAIA and HIPPARCOS data to constrain the dynamical mass of the planet. From an instrumental point of view, these data demonstrate that if they exist, SPHERE can detect young and massive Jupiters as close as 1.6 au from a star located at 20\,pc.

\begin{acknowledgements} 
We acknowledge financial support from the Programme National de Plan\'etologie (PNP) and the Programme National de Physique Stellaire (PNPS) of CNRS-INSU. This work has also been supported by a grant from the French Labex OSUG@2020 (Investissements d'avenir – ANR10 LABX56). The project is supported by CNRS, by the Agence Nationale de la Recherche (ANR-14-CE33-0018). This work has made use of the the SPHERE Data Centre, jointly operated by OSUG/IPAG (Grenoble), PYTHEAS/LAM/CESAM (Marseille), OCA/Lagrange (Nice) and Observatoire de Paris/LESIA (Paris). AML thanks Julian Mejia for his help during the September 2018 DVM observations, Nad\`{e}ge Meunier, Eric Lagadec, and Philippe Delorme for their assistance in the DC process, and Pascal Rubini for his help in the scheduling of the observations. E.S., R.G., D.M., S.D. and R.U.C. acknowledge support from the "Progetti Premiali" funding scheme of the Italian Ministry of Education, University, and Research. SPHERE is an instrument designed and built by a consortium consisting of IPAG (Grenoble,France), MPIA (Heidelberg, Germany), LAM (Marseille, France), LESIA (Paris, France), Laboratoire Lagrange (Nice, France), INAF Osservatorio Astronomico di Padova (Italy), Observatoire de Gen\`{e}ve (Switzerland), ETH Zurich (Switzerland), NOVA (Netherlands), ONERA (France) and ASTRON (Netherlands) in collaboration with ESO. SPHERE was funded by ESO, with additional contributions from CNRS (France), MPIA (Germany), INAF (Italy), FINES (Switzerland) and NOVA (Netherlands). SPHERE also received funding from the European Commission Sixth and Seventh Framework Programmes as part of the Optical Infrared Coordination Network for Astronomy (OPTICON) under grant number RII3-Ct-2004-001566 for FP6 (2004-2008), grant number 226604 for FP7 (2009-2012) and grant number 312430 for FP7 (2013-2016).
\end{acknowledgements}

\bibliographystyle{aa}
\bibliography{biblio}

\begin{thebibliography}{28}
\expandafter\ifx\csname natexlab\endcsname\relax\def\natexlab#1{#1}\fi

\bibitem[{{Beust} \& {Morbidelli}(1996)}]{bmo96}
{Beust}, H. \& {Morbidelli}, A. 1996, \icarus, 120, 358

\bibitem[{{Beust} \& {Morbidelli}(2000)}]{bmo00}
{Beust}, H. \& {Morbidelli}, A. 2000, \icarus, 143, 170

\bibitem[{{Beuzit} {et~al.}(2008){Beuzit}, {Feldt}, {Dohlen}, {Mouillet},
  {Puget}, {Wildi}, {Abe}, {Antichi}, {Baruffolo}, {Baudoz}, {Boccaletti},
  {Carbillet}, {Charton}, {Claudi}, {Downing}, {Fabron}, {Feautrier},
  {Fedrigo}, {Fusco}, {Gach}, {Gratton}, {Henning}, {Hubin}, {Joos}, {Kasper},
  {Langlois}, {Lenzen}, {Moutou}, {Pavlov}, {Petit}, {Pragt}, {Rabou}, {Rigal},
  {Roelfsema}, {Rousset}, {Saisse}, {Schmid}, {Stadler}, {Thalmann}, {Turatto},
  {Udry}, {Vakili}, \& {Waters}}]{beuzit08}
{Beuzit}, J.-L., {Feldt}, M., {Dohlen}, K., {et~al.} 2008, in Society of
  Photo-Optical Instrumentation Engineers (SPIE) Conference Series, Vol. 7014

\bibitem[{{Bonnefoy} {et~al.}(2014{\natexlab{a}}){Bonnefoy}, {Marleau},
  {Galicher}, {Beust}, {Lagrange}, {Baudino}, {Chauvin}, {Borgniet}, {Meunier},
  {Rameau}, {Boccaletti}, {Cumming}, {Helling}, {Homeier}, {Allard}, \&
  {Delorme}}]{bonnefoy14}
{Bonnefoy}, M., {Marleau}, G.-D., {Galicher}, R., {et~al.} 2014{\natexlab{a}},
  \aap, 567, L9

\bibitem[{{Bonnefoy} {et~al.}(2014{\natexlab{b}}){Bonnefoy}, {Marleau},
  {Galicher}, {Beust}, {Lagrange}, {Baudino}, {Chauvin}, {Borgniet}, {Meunier},
  {Rameau}, {Boccaletti}, {Cumming}, {Helling}, {Homeier}, {Allard}, \&
  {Delorme}}]{mbo14}
{Bonnefoy}, M., {Marleau}, G.-D., {Galicher}, R., {et~al.} 2014{\natexlab{b}},
  \aap, 567, L9

\bibitem[{Carbillet {et~al.}(2011)Carbillet, Bendjoya, Abe, Guerri, Boccaletti,
  Daban, Dohlen, Ferrari, Robbe-Dubois, Douet, \& Vakili}]{Carbillet2011}
Carbillet, M., Bendjoya, P., Abe, L., {et~al.} 2011, Experimental Astronomy,
  30, 39

\bibitem[{{Chauvin} {et~al.}(2017{\natexlab{a}}){Chauvin}, {Desidera},
  {Lagrange}, {Vigan}, {Feldt}, {Gratton}, {Langlois}, {Cheetham}, {Bonnefoy},
  \& {Meyer}}]{chauvin17b}
{Chauvin}, G., {Desidera}, S., {Lagrange}, A.-M., {et~al.} 2017{\natexlab{a}},
  in SF2A-2017: Proceedings of the Annual meeting of the French Society of
  Astronomy and Astrophysics, ed. C.~{Reyl{\'e}}, P.~{Di Matteo}, F.~{Herpin},
  E.~{Lagadec}, A.~{Lan{\c c}on}, Z.~{Meliani}, \& F.~{Royer}, 331--335

\bibitem[{{Chauvin} {et~al.}(2017{\natexlab{b}}){Chauvin}, {Desidera},
  {Lagrange}, {Vigan}, {Gratton}, {Langlois}, {Bonnefoy}, {Beuzit}, {Feldt},
  {Mouillet}, {Meyer}, {Cheetham}, {Biller}, {Boccaletti}, {D'Orazi},
  {Galicher}, {Hagelberg}, {Maire}, {Mesa}, {Olofsson}, {Samland}, {Schmidt},
  {Sissa}, {Bonavita}, {Charnay}, {Cudel}, {Daemgen}, {Delorme},
  {Janin-Potiron}, {Janson}, {Keppler}, {Le Coroller}, {Ligi}, {Marleau},
  {Messina}, {Molli{\`e}re}, {Mordasini}, {M{\"u}ller}, {Peretti}, {Perrot},
  {Rodet}, {Rouan}, {Zurlo}, {Dominik}, {Henning}, {Menard}, {Schmid},
  {Turatto}, {Udry}, {Vakili}, {Abe}, {Antichi}, {Baruffolo}, {Baudoz},
  {Baudrand}, {Blanchard}, {Bazzon}, {Buey}, {Carbillet}, {Carle}, {Charton},
  {Cascone}, {Claudi}, {Costille}, {Deboulbe}, {De Caprio}, {Dohlen},
  {Fantinel}, {Feautrier}, {Fusco}, {Gigan}, {Giro}, {Gisler}, {Gluck},
  {Hubin}, {Hugot}, {Jaquet}, {Kasper}, {Madec}, {Magnard}, {Martinez},
  {Maurel}, {Le Mignant}, {M{\"o}ller-Nilsson}, {Llored}, {Moulin},
  {Orign{\'e}}, {Pavlov}, {Perret}, {Petit}, {Pragt}, {Puget}, {Rabou},
  {Ramos}, {Rigal}, {Rochat}, {Roelfsema}, {Rousset}, {Roux}, {Salasnich},
  {Sauvage}, {Sevin}, {Soenke}, {Stadler}, {Suarez}, {Weber}, {Wildi},
  {Antoniucci}, {Augereau}, {Baudino}, {Brandner}, {Engler}, {Girard}, {Gry},
  {Kral}, {Kopytova}, {Lagadec}, {Milli}, {Moutou}, {Schlieder},
  {Szul{\'a}gyi}, {Thalmann}, \& {Wahhaj}}]{chauvin17}
{Chauvin}, G., {Desidera}, S., {Lagrange}, A.-M., {et~al.} 2017{\natexlab{b}},
  \aap, 605, L9

\bibitem[{{Chauvin} {et~al.}(2012){Chauvin}, {Lagrange}, {Beust}, {Bonnefoy},
  {Boccaletti}, {Apai}, {Allard}, {Ehrenreich}, {Girard}, {Mouillet}, \&
  {Rouan}}]{chauvin12}
{Chauvin}, G., {Lagrange}, A.-M., {Beust}, H., {et~al.} 2012, \aap, 542, A41

\bibitem[{{Claudi} {et~al.}(2008){Claudi}, {Turatto}, {Gratton}, {Antichi},
  {Bonavita}, {Bruno}, {Cascone}, {De Caprio}, {Desidera}, {Giro}, {Mesa},
  {Scuderi}, {Dohlen}, {Beuzit}, \& {Puget}}]{claudi08}
{Claudi}, R.~U., {Turatto}, M., {Gratton}, R.~G., {et~al.} 2008, in Society of
  Photo-Optical Instrumentation Engineers (SPIE) Conference Series, Vol. 7014,
  Society of Photo-Optical Instrumentation Engineers (SPIE) Conference Series,
  3

\bibitem[{{Dohlen} {et~al.}(2008){Dohlen}, {Langlois}, {Saisse}, {Hill},
  {Origne}, {Jacquet}, {Fabron}, {Blanc}, {Llored}, {Carle}, {Moutou}, {Vigan},
  {Boccaletti}, {Carbillet}, {Mouillet}, \& {Beuzit}}]{dohlen08}
{Dohlen}, K., {Langlois}, M., {Saisse}, M., {et~al.} 2008, in Society of
  Photo-Optical Instrumentation Engineers (SPIE) Conference Series, Vol. 7014,
  Society of Photo-Optical Instrumentation Engineers (SPIE) Conference Series,
  3

\bibitem[{{Galicher} {et~al.}(2018){Galicher}, {Boccaletti}, {Mesa}, {Delorme},
  {Gratton}, {Langlois}, {Lagrange}, {Maire}, {Le Coroller}, {Chauvin},
  {Biller}, {Cantalloube}, {Janson}, {Lagadec}, {Meunier}, {Vigan},
  {Hagelberg}, {Bonnefoy}, {Zurlo}, {Rocha}, {Maurel}, {Jaquet}, {Buey}, \&
  {Weber}}]{galicher18}
{Galicher}, R., {Boccaletti}, A., {Mesa}, D., {et~al.} 2018, \aap, 615, A92

\bibitem[{{Kiefer} {et~al.}(2014){Kiefer}, {Lecavelier des Etangs}, {Boissier},
  {Vidal-Madjar}, {Beust}, {Lagrange}, {H{\'e}brard}, \& {Ferlet}}]{kiefer14}
{Kiefer}, F., {Lecavelier des Etangs}, A., {Boissier}, J., {et~al.} 2014, \nat,
  514, 462

\bibitem[{{Lagrange} {et~al.}(2012{\natexlab{a}}){Lagrange}, {Boccaletti},
  {Milli}, {Chauvin}, {Bonnefoy}, {Mouillet}, {Augereau}, {Girard}, {Lacour},
  \& {Apai}}]{lagrange12disk}
{Lagrange}, A.-M., {Boccaletti}, A., {Milli}, J., {et~al.} 2012{\natexlab{a}},
  \aap, 542, A40

\bibitem[{{Lagrange} {et~al.}(2010){Lagrange}, {Bonnefoy}, {Chauvin}, {Apai},
  {Ehrenreich}, {Boccaletti}, {Gratadour}, {Rouan}, {Mouillet}, {Lacour}, \&
  {Kasper}}]{lagrange10}
{Lagrange}, A.-M., {Bonnefoy}, M., {Chauvin}, G., {et~al.} 2010, Science, 329,
  57

\bibitem[{{Lagrange} {et~al.}(2012{\natexlab{b}}){Lagrange}, {De Bondt},
  {Meunier}, {Sterzik}, {Beust}, \& {Galland}}]{lagrange12rv}
{Lagrange}, A.-M., {De Bondt}, K., {Meunier}, N., {et~al.} 2012{\natexlab{b}},
  \aap, 542, A18

\bibitem[{{Lecavelier Des Etangs} {et~al.}(1995){Lecavelier Des Etangs},
  {Deleuil}, {Vidal-Madjar}, {Ferlet}, {Nitschelm}, {Nicolet}, \&
  {Lagrange-Henri}}]{leca95}
{Lecavelier Des Etangs}, A., {Deleuil}, M., {Vidal-Madjar}, A., {et~al.} 1995,
  \aap, 299, 557

\bibitem[{{Lecavelier des Etangs} \& {Vidal-Madjar}(2016)}]{leca16}
{Lecavelier des Etangs}, A. \& {Vidal-Madjar}, A. 2016, \aap, 588, A60

\bibitem[{{Lecavelier Des Etangs} {et~al.}(1997){Lecavelier Des Etangs},
  {Vidal-Madjar}, {Burki}, {Lamers}, {Ferlet}, {Nitschelm}, \&
  {Sevre}}]{leca97}
{Lecavelier Des Etangs}, A., {Vidal-Madjar}, A., {Burki}, G., {et~al.} 1997,
  \aap, 328, 311

\bibitem[{{Maire} {et~al.}(2016){Maire}, {Langlois}, {Dohlen}, {Lagrange},
  {Gratton}, {Chauvin}, {Desidera}, {Girard}, {Milli}, {Vigan}, {Zins},
  {Delorme}, {Beuzit}, {Claudi}, {Feldt}, {Mouillet}, {Puget}, {Turatto}, \&
  {Wildi}}]{maire16}
{Maire}, A.-L., {Langlois}, M., {Dohlen}, K., {et~al.} 2016, in \procspie, Vol.
  9908, Ground-based and Airborne Instrumentation for Astronomy VI, 990834

\bibitem[{{Marois} {et~al.}(2014){Marois}, {Correia}, {V{\'e}ran}, \&
  {Currie}}]{Marois14}
{Marois}, C., {Correia}, C., {V{\'e}ran}, J.-P., \& {Currie}, T. 2014, in IAU
  Symposium, Vol. 299, IAU Symposium, ed. M.~{Booth}, B.~C. {Matthews}, \&
  J.~R. {Graham}, 48--49

\bibitem[{{Marois} {et~al.}(2006){Marois}, {Lafreni{\`e}re}, {Doyon},
  {Macintosh}, \& {Nadeau}}]{marois06}
{Marois}, C., {Lafreni{\`e}re}, D., {Doyon}, R., {Macintosh}, B., \& {Nadeau},
  D. 2006, \apj, 641, 556

\bibitem[{{Smith} \& {Terrile}(1984)}]{smith84}
{Smith}, B.~A. \& {Terrile}, R.~J. 1984, Science, 226, 1421

\bibitem[{{Snellen} \& {Brown}(2018)}]{snellen18}
{Snellen}, I.~A.~G. \& {Brown}, A.~G.~A. 2018, Nature Astronomy

\bibitem[{{Soummer} {et~al.}(2012){Soummer}, {Pueyo}, \& {Larkin}}]{soummer12}
{Soummer}, R., {Pueyo}, L., \& {Larkin}, J. 2012, \apjl, 755, L28

\bibitem[{{Th{\'e}bault} \& {Beust}(2001)}]{thebault01}
{Th{\'e}bault}, P. \& {Beust}, H. 2001, \aap, 376, 621

\bibitem[{{van Leeuwen}(2007)}]{vanLeeuwen07}
{van Leeuwen}, F. 2007, \aap, 474, 653

\bibitem[{{Wang} {et~al.}(2016){Wang}, {Graham}, {Pueyo}, {Kalas},
  {Millar-Blanchaer}, {Ruffio}, {De Rosa}, {Ammons}, {Arriaga}, {Bailey},
  {Barman}, {Bulger}, {Burrows}, {Cardwell}, {Chen}, {Chilcote}, {Cotten},
  {Fitzgerald}, {Follette}, {Doyon}, {Duch{\^e}ne}, {Greenbaum}, {Hibon},
  {Hung}, {Ingraham}, {Konopacky}, {Larkin}, {Macintosh}, {Maire}, {Marchis},
  {Marley}, {Marois}, {Metchev}, {Nielsen}, {Oppenheimer}, {Palmer}, {Patel},
  {Patience}, {Perrin}, {Poyneer}, {Rajan}, {Rameau}, {Rantakyr{\"o}},
  {Savransky}, {Sivaramakrishnan}, {Song}, {Soummer}, {Thomas}, {Vasisht},
  {Vega}, {Wallace}, {Ward-Duong}, {Wiktorowicz}, \& {Wolff}}]{wang16}
{Wang}, J.~J., {Graham}, J.~R., {Pueyo}, L., {et~al.} 2016, \aj, 152, 97

\end{thebibliography}

\end{document}